# Evidence for undamped waves on ohmic materials

J. M. A. Figueiredo[*] and Roberto B. Sardenberg[†]
Unversidade Federal de Minas Gerais - Departamento de Física
Caixa Postal 702 - 30.123-970 - Belo Horioznte - Brazil

## Abstract

The propagation of electromagnetic fields in matter has been the subject of intensive studies since the discovery of its rich dynamics. Impedance measurements are one most used technique available to study material properties as well as electromagnetic devices and circuits. This way, novelties on device construction and circuit technology associated to new material properties and/or unusual field dynamics generally rely on results supported by impedance data. Recent advances on nanostructured materials explore astounding molecular properties derived from nanoscale levels and apply them to studies foucused on the generation of new devices. Accordingly, properties inherent to quantum dynamics can also generate unusual circuit elements, not included on the former development of the electromagnetic theory. On same footings, advances in field dynamics could also determine the advent of new technologies, producing immediate impact on our everyday life. In this work we present the results obtained by measuring the impedance of single spires and coils of specific geometry in the MHz range. They demonstrate that a new passive circuit element was found, which bears out the existence of an as yet unobserved propagation mode of the electromagnetic fields in matter. Our results also indicates that this effect is more evident using carbon made spires.

The propagation of electromagnetic fields in matter has been the subject of intensive studies since the discovery of its rich dynamics. The variety of material properties relevant to the interaction of these fields with matter led to the development of passive circuit elements and active devices designed to control charges and currents in matter. This allows the advent of technologies inherent to almost all aspects of modern human life. Passive circuit elements are based on equations between field and matter variables, as defined by the constitutive relations, and are crucial in determining the best performance and economic viability of practical circuits. In this case, a property of Maxwell equations which has not yet been considered could determine new relations between fields and currents, thus allowing new circuit configurations or even new circuit elements.
In this work we present the results obtained by measuring the impedance of single spires and coils of specific geometry in the MHz range. They demonstrate that a new passive circuit element was found, which bears out the existence of an as yet unobserved propagation mode of the electromagnetic fields in matter.
Due to the immediate applicability of almost all electromagnetic phenomena, materials containing non-trivial electronic properties are highly sought-after. To this end, there has recently been a marked search for new materials or special molecular arrangements in the nanoscale domain, featuring better device performance, low fabrication cost and/or low energy consumption [1], [2]. On more general grounds, Strukov, Snider, Stewart and Williams [3] recently found the menristor, the last circuit element that can be derived from the fundamental relations of circuit variables, which was predicted by Chua[4]. Therefore, any novelty in circuit configuration or circuit size, such as the use of quantum dynamics in practical devices [5], [6] or new materials [7], [8], [9], [10] can enhance their applicability. Likewise, the discovery of new field dynamics in matter could be of relevance in determining properties of existing circuit elements or even lead to the development of new ones.

Here we report our results of impedance measurements on ohmic materials. These measurements were performed in the cylindrical geometry (wires), by using AC fields in the kHz to MHz frequency range. Due to the inherent dissipative character of the ohmic regime, it is expected that the real part of the impedance increases as the frequency of the applied excitation signal becomes larger. It is known that this effect is caused by magnetic forces inside the material, which push charge carriers to the longitudinal surface of the material in such a way that the effective transversal area available to charge transport becomes smaller than the geometric one - the skin effect. However, for longer wires we have observed that this effect is lessened, being almost minimal. Our data show that, as frequency increases the real part of the impedance gets a maximum. Further increases in frequency lead to smaller values of the real part, which may even assume negative values. This is an effect that field dynamics relevant to skin calculations cannot account for. Then for still larger frequencies a kind of aberrant resonance peak, in the sense that it is inverted when compared to usual RLC peaks, is observed. This means that the spire presents a "negative resistance" component forming its impedance spectral response. As a consequence, the observed resonances admit an extended phase angle variation, ranging from $-\pi$ to $\pi$ instead of the usual $-\pi/2$ to $\pi/2$ interval, defined in its limits by the reactance of pure capacitances and inductances.

Moreover, the observed impedance profile may change, depending on the wire material and spire topology. In order to get an insight as to which kind of field dynamics may cause this effect, we also present a simple theoretical model that accounts, at least qualitatively, for the major aspects of the data presented here. Our calculation shows that the impedance of single spires seems to derive from interference of undumped waves propagating along the wire.

## Measurements and experimental data

Measurements were performed on single spires and on coils using a Novocontrol Alpha-A High Performance Frequency Analyzer Impedancemeter. It has a $10$ TΩ input resistance in parallel with a $0.5$ pF input capacitance. Its frequency range spans the μHz to 40MHz interval. Spires were made of Al, Cu, NiFe alloy and carbon fiber. Coils were made of Al, Cu, brass and constantan. Measurements were performed in the frequency range of 10kHz to 35MHz and with an excitation amplitude of 50mV. Qualitative measurements in the frequency range of 10 to 100MHz were also performed, using an external RF signal generator (Rohde & Schwarz - SMY01) and a 100MHz dual channel oscilloscope (Tektronix TDS 220). In this case, one channel reads the applied voltage in a coil and the other the voltage on a small piece of a pure resistive material placed in series with the coil and having an almost zero degree phase angle. This way, the voltage in the second channel is in phase with the circuit current. This type of measurement generates less precise values for the impedance when compared to the impedancemeter but provides a direct comparison of voltage and current in the coil as function of time.

Two kinds of circuits were assembled. One type consisted of large single loop spires. These spires were built in rectangular (60 to 120mm wide) geometry. The metallic ones had a wire diameter ranging from 10 to 1000μm and wire length from 4 to 26m. Carbon fiber spires were also made. This fiber is a bundle of $\sim 5$ μm wires forming a 100μm thick sheet. These spires were made with a $\sim 1$ mm wide and 4m long sheet or $\sim 2$ mm wide and 2m long sheet. All of them were rectangular and made with a wire separation within the 40mm to 130mm range. In this work, spires are designed by use of the format material/wire length/wire separation/wire diameter.

The other circuit type consisted of coils having a wire length around 15m and coil diameter of 40mm. In order to avoid relevant inter-spire capacitance, which may drive the phase angle to $-\pi/2$ at high frequencies, the spires and coils were made with enough wire separation distance which, in the coils, ranged from 2 to 5mm. However, spire area in both kinds of circuits had to be controlled as well to prevent external inductances from driving the phase angle to $\pi/2$. This way, the best compromise for wire separation and circuit area was empirically determined, which maximized the measured negative real part impedance. Coils are designed by use of the format material/wire length/coil separation/wire diameter.

## Spires measurements

For an excitation signal of a few kilohertz the impedance of spires presents, as expected, a characteristic RL response. Moreover, due to skin effects, an additional increase of the impedance as a function of the frequency was also observed. However, as shown in Figure 1(a), for frequencies of few Megahertz and above, the real part of the impedance becomes smaller and assumes negative values. Once the real part is negative, the third and fourth quadrants of the phase angle is attained. This new effect appears to be more intense in long wires (quite a few meters in the MHz region) having small cross-sectional area (dozen microns diameter). Increments in the frequency leads to a kind of resonant figure, where both, the real and imaginary parts present points of extrema in their spectral profile. Spectra for the real part are shown on Figure 1(a), where the presence of a negative peak is evident.

Metallic spires present negative peaks ranging from a few to dozens kohms. Among them, the skin effect was more evident on the NiFe spire due to its high magnetic susceptibility. In the carbon fiber spire case, a negative peak of $-488$ kΩ was observed, using a wire length of only 4m, smaller than those used in all metallic spires. We also show in Figure 1(b) a separate plot for this spire presenting its real part impedance and its calculated phase angle. The transition from $\pi$ to $-\pi$ (the crossing of the negative real axis) is evident. Having passed the resonance, the impedance tends to assume its regular profile, where its real part assumes values in the range expected for DC or skin figures. We show in Figure 2 the complete resonance profile for another carbon spire.

## Coils measurements

For coils, the challenge of controlling spare effects of both inter-spire capacitance and external inductance is harder. Since long wire length is crucial in detecting negative resistance effects, a large number of spires in a coil is also demanded. Thus a good balance of wire length, inter-spire gap and coil diameter is necessary in order to get the best results. Once again, this compromise was empirically obtained. The spectra we measured for coils are quite different from those observed in the single spires, the main difference being a larger number of inverted peaks in coils, but not too deep when compared to single spires. This effect is shown in Figure 3, which presents data for a single aluminum spire and for a constantan coil. It appears that coils of more resistive materials present deeper negative peaks even though the inter-spire capacitance could limit direct observation of the effect. In fact, multiple deep peaks were obtained using this constantan coil. Another relevant feature presented in Figure 3 is a qualitative disagreement between the spectral profile of metallic and carbon spires. We have verified that in all metallic spires the imaginary part spectrum, at frequencies close to the inverted peak, reaches its negative minimum first and then rises to a positive maximum. The analogue spectrum for carbon spires initially presents a strong positive spike before its transition to negative values. In short, the signal of the derivative at the $Z_I = 0$ point is opposite for metallic and carbon spires. This effect can be checked by comparing Figures 2 and 3.

Measurements using an oscilloscope were performed on coils. This experiment allows a view of negative real-part impedances if an in-series, pure resistive, small value shunt is used as a current sensor. In this case, current in the circuit is in-phase with the voltage in the shunt in such a way that a direct comparison of voltage and current in the coil, as function of time, is possible. In order to achieve a good experimental condition an external RF generator was linked to a coil and to a 30Ω shunt. Channel 1 of the oscilloscope monitors the applied signal and channel 2 the voltage in the shunt. Figure 4(a) shows oscilloscope data for measurements made at 17MHz using an aluminum coil. Comparing both channels we can see a characteristic anti-phase picture, directly confirming the negative real part impedance. Extreme anti-phasing, as shown here, must happen at a frequency where a negative real part coincides with a null imaginary part. This can be checked in Figure 4(b), where the spectrum of this coil is shown, confirming a change in the signal of the imaginary part close to 17MHz. The real part assumes negative values, in total agreement with the oscilloscope data. Detail of this spectral region is shown in Figure 4(c). On an oscilloscope screen, visual confirmation of the anti-phase pattern in the 100 to 500MHz frequency range was also verified, with the use of a similar circuit configuration. It consisted of a linear wire made of carbon fiber, having a length of 0.18m and in series with a 5mm length shunt made of the same material. Multiple inverted peaks were observed at various frequency values within this interval.

## Data analysis

The inverted resonance peak observed in spires strongly resembles the spectral response of a lumped element circuit. In this sense, we adjusted spectral data in the inverted peak region for the (C/2m/10cm/1mm) spire, to the impedance of a parallel RLC circuit according to the formula $Z(\omega) = R_0 + [R^{-1} + i\omega C + (i\omega L)^{-1}]^{-1}$. As shown in Figure 2, this formula is well suited to the real and imaginary parts of our data. Adjusted parameters are $R_0 = (1.12 \pm 0.07) \cdot 10^3 \Omega$, $R = -(2.29 \pm 0.03) \cdot 10^4 \Omega$, $C = (1.27 \pm 0.02) \cdot 10^{-12} F$, $L = (1.86 \pm 0.03) \cdot 10^{-6} H$. The main result of our fitting is the existence of a circuit element presenting a negative resistance character. Its presence drives the impedance angle to the third and fourth quadrants and can, therefore, be characterized as a new circuit element. Differently from an ohmic resistance, where applied electric field and current density vectors are parallel, here they are anti-parallel at the source inputs. As shown bellow, this fact corroborates the existence of a field pattern along the wire.

For metallic spires and coils, it appears that material properties can affect the real part of the impedance profile mainly in two ways: either by adjusting the (inverted peak) resonant frequency or by defining the crossover for the transition of regular (Zr>0) to negative impedance values. The impressive negative impedance values measured for the carbon spire indicate that transport properties may also be relevant in determining the spectral response. The imaginary part seems to be very sensitive to spire geometry because stray capacitances and inductances can easily change its profile. In short, whatever the used geometry, the general profile of an inverted peak was present everywhen a negative real part was attained.

We get a better insight concerning the origin of the inverted peak by using a model of the impedance in a spire assuming that undamped waves are propagating along the wire. This is not a simple hypothesis because dissipation is always present in an ohmic material. Anyway, consider that a pattern for the potential difference in the wire (relative to z=0) exists and is given by: $V(z,t) = [V_+ \exp(iqz) + V_- \exp(-iqz)] \exp(i\omega t)$. The electric field derived from this potential is given by $E = -\frac{dV}{dz}$ and the current density by $\sigma E$. The vector potential contribution is not considered here, but could also be important, which is discussed below. Electric current associated to this field is given by $I = [I_+ \exp(iqz) - I_- \exp(-iqz)] \exp(i\omega t)$ and the impedance is then calculated as the ratio of voltage and current existing at $z = l$, the wire's lenght. That is

$$Z(\omega) = \frac{V(l,t)}{I(l,t)} = \frac{V_+(\omega)\exp(iql) + V_-(\omega)\exp(-iql)}{I_+(\omega)\exp(iql) - I_-(\omega)\exp(-iql)}$$

$$Z(\omega) = Z_0(\omega) \frac{1 + \rho(\omega)\exp(-2iql)}{1 - \rho(\omega)\exp(-2iql)}$$

where $Z_0(\omega) \equiv \frac{V_+(\omega)}{I_+(\omega)}$ is a characteristic impedance and $\rho(\omega) \equiv \frac{V_-(\omega)}{V_+(\omega)}$ the reflection coefficient. This reasoning is similar to those used in transmission line impedance calculations. Then we can write

$$Z(\omega) = Z_0(\omega)\frac{[1+\rho(\omega)\exp(-2iql)][1-\bar{\rho}(\omega)\exp(2iql)]}{|1-\rho(\omega)\exp(-2iql)|^2}$$

Now by assuming that the characteristic impedance is real (for example, its dc resistance), we get the impedance proportional to

$$Z(\omega) \simeq \frac{1-|\rho(\omega)|^2 - 2i\,\text{Im}(\rho(\omega)\exp(-2iql))}{1+|\rho(\omega)|^2}$$

The real part of this expression becomes negative if the reflected waves dominate above some frequency. We made guesses for the reflection coefficient and got a profile similar to our data when $\rho(\omega)$ and $q(\omega)$ are both proportional to $\omega$. In fact, a careful choice of numerical values for these functions leads to an expression presenting a qualitative profile like that of our data. That is, assuming a linear dependence with the frequency, we got an expression of the type:

$$Z(x) \simeq \frac{\left[1+0.2x\exp\left(i(\frac{\pi}{4}+x)b\right)\right]\left[1-0.2x\exp\left(-i(\frac{\pi}{4}+x)b\right)\right]}{\left|1-0.2x\exp\left(i(\frac{\pi}{4}+x)b\right)\right|^2}$$

and a plot is given on Figure 5. It displays a profile presenting all relevant features observed on our impedance data, including the inversion of the imaginary part profile observed on metallic and carbon spires.

The way these waves may propagate without attenuation in an ohmic material is not trivial. However, an analysis of the wave equation for the electric field inside the wire may give some clues. In fact, in the cylindrical geometry the wave equation for the electric field is given by

$$\nabla_r^2 E_z(r,z) + \frac{d^2 E_z}{dr^2} + \frac{\omega^2}{c^2}E_z = i\omega\mu\sigma E_z,$$

where $\nabla_r^2$ is the radial part of the laplacian. Assuming that $E_z = f(r)\exp(iqz)$ we get $\nabla_r^2 f(r) = i\omega\mu\sigma f$ and $q = \frac{\omega}{c}$. Solution of the first equation is a zero order Bessel function. The second one leads to a real-valued q, as demanded for undamped traveling waves. Thus, undamped spatial modes for the longitudinal electric field may exist if the three-dimensional character of the wire is considered in the wave propagation process. It seems that dissipation could occur only in the radial prefactor (the usual skin effect), leaving the longitudinal direction free for undamped propagation.

However, the problem is not so straightforward and, as it was stated, has not been totally unraveled yet. The magnetic field associated to $E_z$ points to the $\hat{\varphi}$ direction. Once we apply Faraday's law we can see that a radial electric field will be generated by this magnetic field as well. So a complex pattern for the current density arises as a result of the interference of the radial and longitudinal fields in such a way that undamped longitudinal waves must survive. This means that contributions derived from the vector potential may also be relevant to this problem. In view of these considerations, a more detailed theoretical analysis of this system wiil be necessary to account for such a complex field dynamics.

## Conclusion

The result of our experiments concerns the observation of unusual impedance values of spires and coils constructed with long wires. At high frequencies the skin effect does not dominate the spectral response anymore and inverse resonance peaks were observed. This effect characterizes the impedance of our circuits as having a negative resistance element. It was explained as the consequence of new field dynamics in material media. Evidence is presented that the existence of undamped waves along the wire is the cause of the negative resistance. We have also presented arguments indicating that these undamped waves can be supported by radial (dissipative) waves in the wire. Since the theory presented here is incomplete, it still requires a more detailed model of this phenomenon .

Negative resistance components can be applied in the construction of extended phase oscilators capable of phase modulation in the whole impedance phase space. A circuit with an extended phase could, in principle, almost double channel capacity in communication systems. Moreover, compact topology of spires and coils can be achieved in the GHz spectral region because much smaller wire length will be necessary to achieve a negative resistance component. This way, projects of circuits working in the extended phase region and designed for practical uses will become feasible.

This work was supported by the Brazilian agencies CNPq/FAPEMIG (INCT) and CAPES.

[*] josef@fisica.ufmg.br
[†] robertobatsar@hotmail.com

**Figures:**

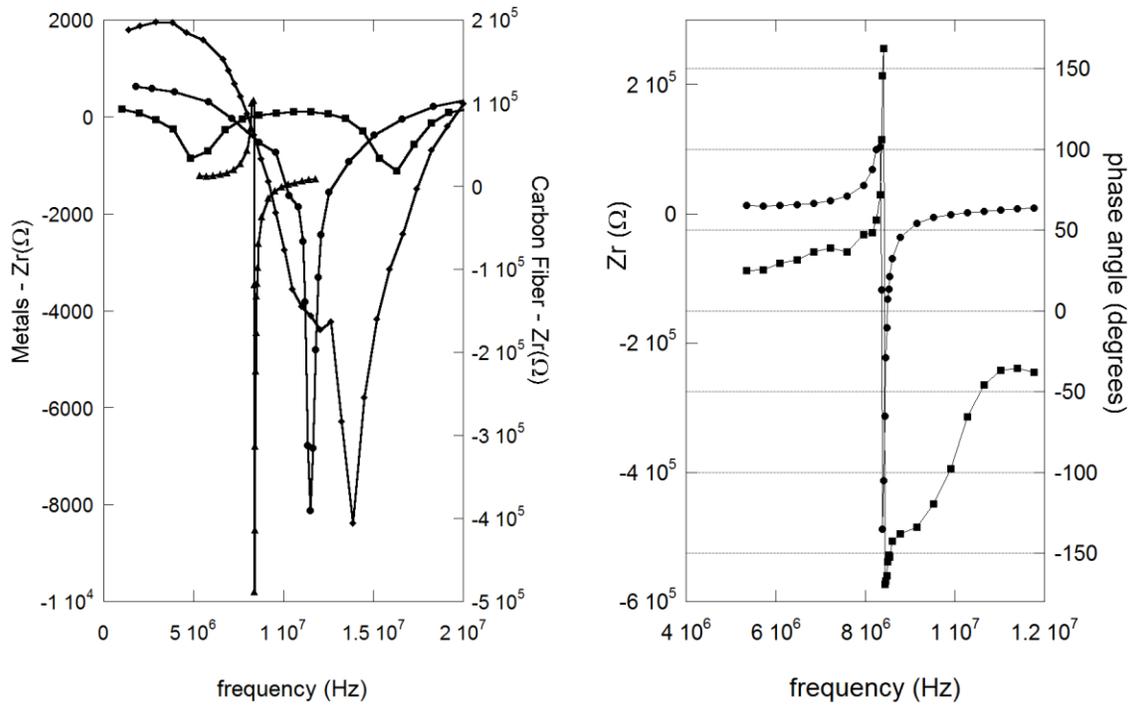

1.a - Real part spectra data of single spires. Left vertical axis corresponds to metallic spires: Al /12$m$/15$cm$/10$\mu m$ ( ● ); Cu/ 19$m$/10$cm$/70$\mu m$ ( ■ ); Ni-Fe/ 5$m$/10$cm$/30$\mu m$ ( ♦ ). Right vertical axis: C/ 4$m$/10$cm$/2$mm$ ( ▲ ). 1.b - Carbon fiber spectrum and phase angle: real part data ( ● ); calculated phase angle ( ■ ).

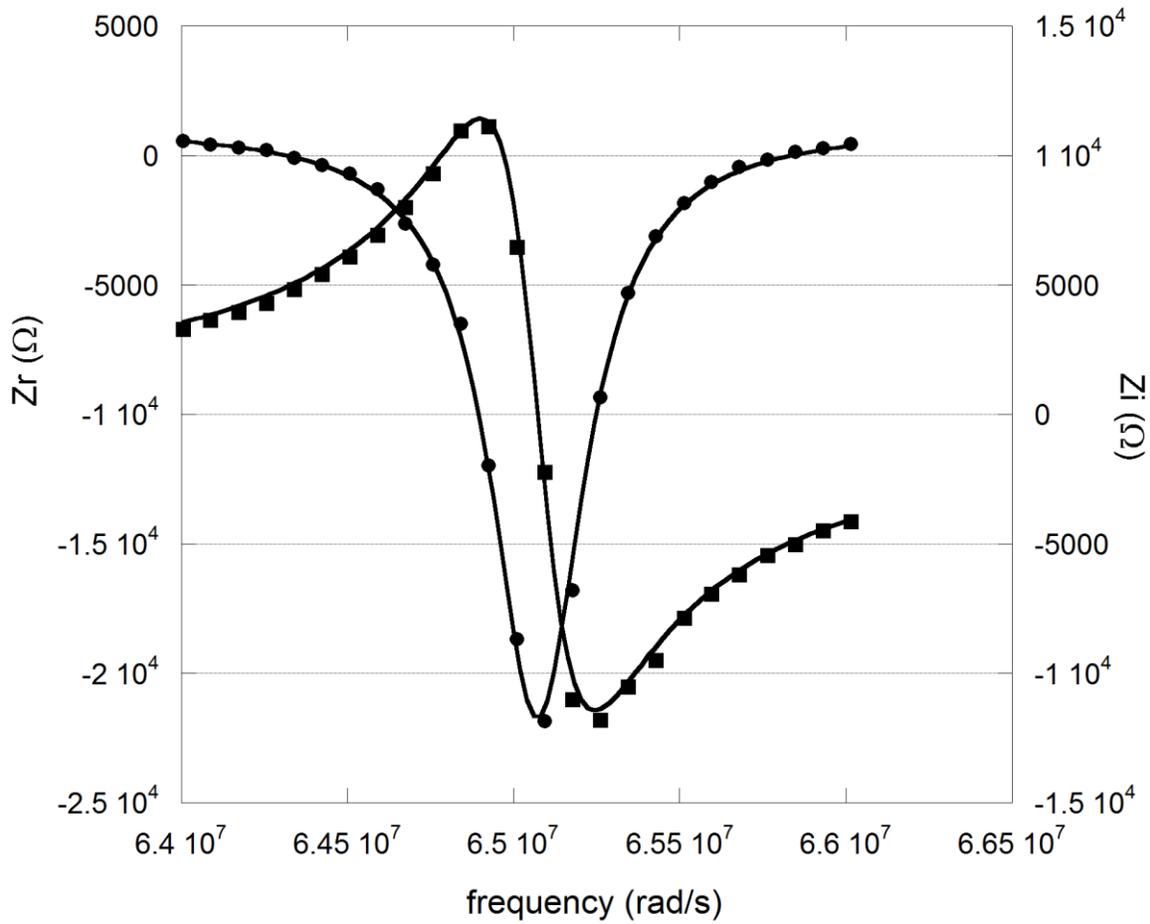

2 - Full spectrum of a carbon fiber spire (C/ $2m/10cm/1mm$ ). Left vertical axis: real part
(●); right vertical axis: imaginary part (■). Solid lines represent a parallel RLC circuit fit.

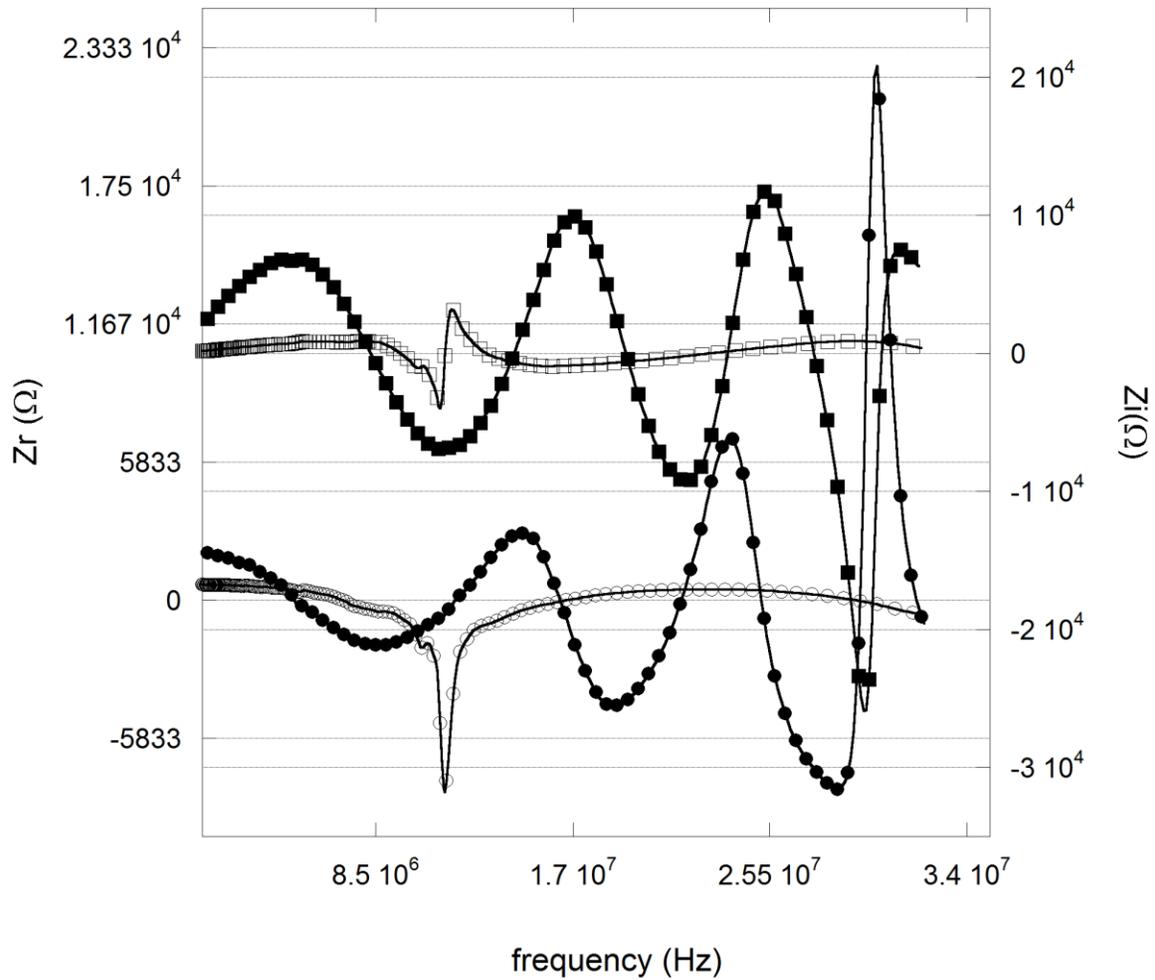

3 - Spectra of a single spire (Al/ $12m/15cm/10\mu m$ ): real part ( ○ ), imaginary part ( □ );
spectra of a coil (constantan/ $27m/1.3mm/100\mu m$ ): real part ( ● ), imaginary part ( ■ ).

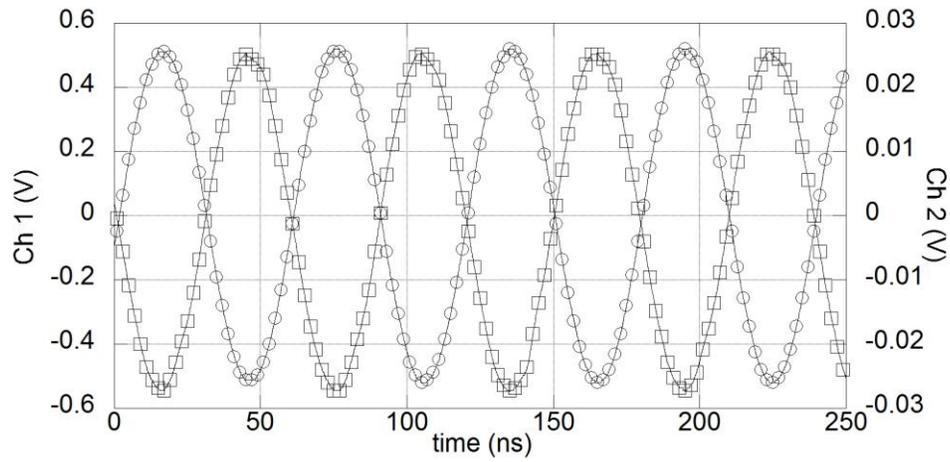
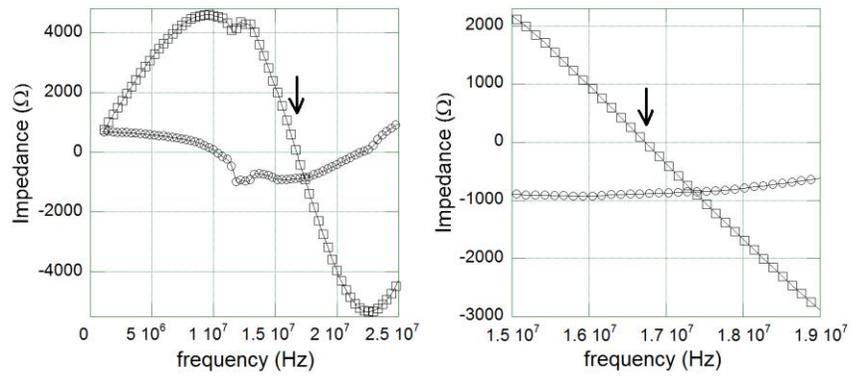

4a - Oscilloscope data for measurements at $17 MHz$. Circles: channel 1; squares: channel 2 voltage. 4b - Full spectrum data: real part ( ○ ); imaginary part ( □ ). An arrow indicates a frequency near $17 MHz$ (the crossing region). 4c - Detail view of the crossing region.

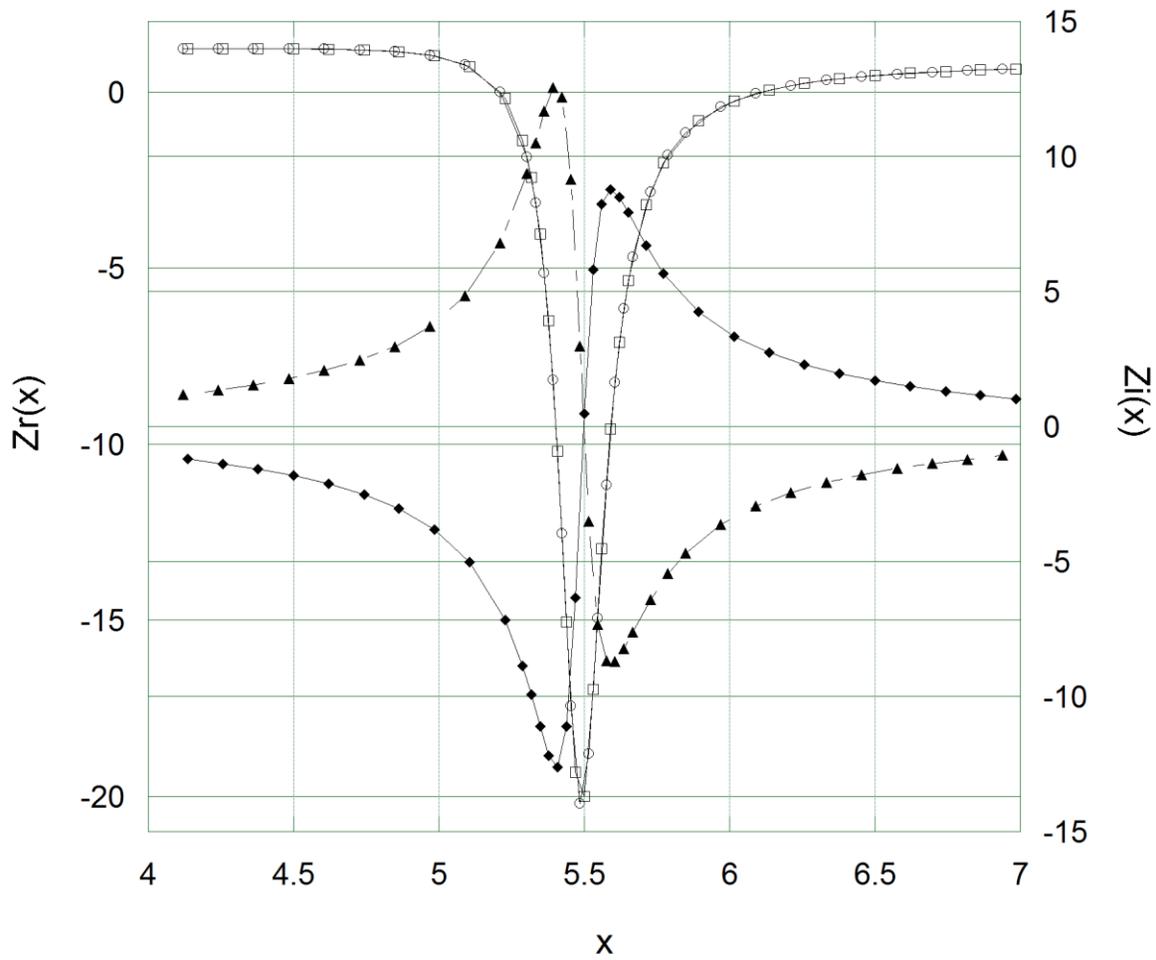

5 - Calculated impedance of a wire assuming that undamped travelling waves are present. Real part, $b = 1$ ( ○ ) ; imaginary part, $b = 1$ ( ♦ ) ; real part, $b = -1$ ( □ ) ; imaginary part, $b = -1$ ( ▲ ). The imaginary part qualitatively displays an inversion in the spectra, also observed on comparing metalic and carbon spires data.